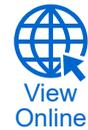
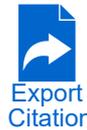

# Multiple Manifestations of Negative Partial Density of States

Kanchan Meena[a)] and P. S. Deo[b)]

S. N. Bose National Centre for Basic Sciences, JD Block, Salt Lake City,700106, Kolkata, India

[a)]Corresponding author: 1996.kanchanmeena@gmail.com
[b)]singhadeoprosenjit@gmail.com

**Astract.** We revealed that with the measurement of the scattering phase shift of electron in low-dimensional or mesoscopic systems local objects of hierarchy of density of states can also determine experimentally. In recent times, it has been exhibited that in mesoscopic systems certain objects of density of states (DOS) hierarchy like local partial DOS, partial DOS, injectivity, emissivity, etc. can become negative in presence of Fano resonance. Negativity of local partial density of states can be interpreted as the losing coherent electrons in reverse time. This may have implications for the thermodynamic properties of these mesoscopic systems. In these negative local partial states, electrons may behave akin to positrons, resulting in practical the possibility of electron-electron interaction. The objective of this research is to reveal some manifestations of local objects in mesoscopic systems, employing rigorous calculations utilizing two different approaches: a continuum model and a discrete or tight binding model. It has been demonstrated that negative local partial states are correlated with Fano-resonance featuring a π phase drop.

## INTRODUCTION

Remarkable advancements in sample fabrication techniques have resulted in allowing us to experimentally access the intermediate regime between classical and quantum physics. The physics of this intermediate regime is generally called mesoscopic physics. We understand classical physics in terms of local variables and objects while quantum mechanics is non-local. In Larmor clock theory [1], Larmor precession of an electron in a magnetic field lead to a concept of local partial density of states (LPDOS) in quantum systems like mesoscopic systems but there were many conceptual problems related to the fact that LPDOS can become negative [2,3] and local objects in quantum systems. These problems have been solved recently, and LPDOS was put on firm mathematical ground using functional analysis from which a clear understanding of time travel emerges that is consistent with quantum mechanics as well as relativity. The purpose of this present work is to reveal theoretical consistencies and find experimental manifestations of negative LPDOS and therefore indirectly, time travel [2,3,4]. Quite counter-intitutively LPDOS ($\rho_{lpd}$) is a local object defined with respect to the leads (in open mesoscopic systems) and help us derive a hierarchy of DOS like partial density of states ($\rho_{pd}$), injectivity ($\rho_i$), emissivity ($\rho_e$), etc. [1,2,3]. Understanding these objects is very crucial to understand mesoscopic response. To calculate these quantities, one has to solve a quantum mechanical electron scattering problem to compute the scattering amplitude as well as the scattering phase shift. It has been seen that the resonances in the mesoscopic systems connected to leads will be mostly Fano resonances. That scattering phase shifts at Fano resonances is very special leading to this rich diversity of physical phenomenon, discussed in the previous paragraph. Studies so far have remained limited to continuum model while the tight binding model (TBM) helps in solving a discretized version of the original problem. If TBM captures some aspects of the phenomena originating from LPDOS then it may become much simpler to study them using the TBM. Such an effort will be also made in this work. We will talk about two possible manifestations. One way to find experimental manifestations of these local objects is through three probe conductance, four probe conductance, etc. We specifically focus on three probe conductance setup and that to only the current in three probe setup. Another experimental manifestation is negative LPDOS may also lead to electron-electron interaction.

# THREE PROBE CONDUCTANCE SETUP

In condensed matter systems to separate contact resistance and sample resistance we use four probe measurement. Such separation is not possible in mesoscopic systems where experimental observations can be interpreted as quantum measurement problem wherein what we measure depends on how we measure. We have shown in Fig. 1 (see details in figure caption) schematic diagram of an experimental setup, which is widely recognized, extensively studied both theoretically and experimentally, and known as the Landauer-Buttiker three probe conductance setup [1,5]. In this setup, lead β makes contact to the sample but does not carry any current to or from the sample due to the adjustment of the chemical potential $\mu_\beta$. The three probe conductance G is given as

$$G = \frac{I_{\alpha\gamma}}{\mu_\beta} = -G_{\alpha\gamma} - \frac{G_{\alpha\beta} G_{\beta\gamma}}{G_{\beta\alpha}+G_{\beta\gamma}} \qquad (1)$$

Here, $I_{\alpha\gamma}$ is three probe current from lead $\gamma$ to $\alpha$ and $\mu_\beta$ is the chemical potential as shown in Fig. 1. $G_{\alpha\gamma} = \frac{2e^2}{h}|S_{\alpha\gamma}|^2$ is two probe conductance, where $|S_{\alpha\gamma}|^2$ is quantum mechanical scattering probability. Likewise, $G_{\alpha\beta}$ and $G_{\beta\gamma}$ can also be defined.

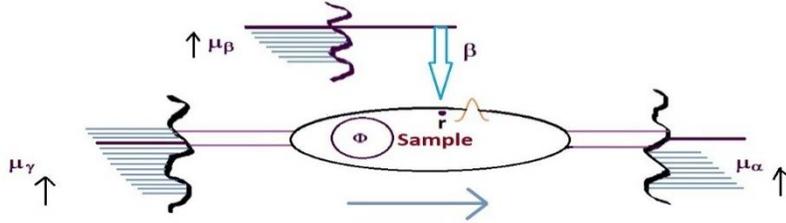

Figure 1. A schematic diagram of three probe conductance setup wherein left reservoir, as a source, has higher chemical potential $\mu_\gamma$ and right reservoir, as a sink, has lower chemical potential $\mu_\alpha$. The current is going from source to sink. Left lead γ and right lead α are attached to the sample wherein Aharonov-Bohm flux Φ within circle as shown. Lead β has equilibrium chemical potential $\mu_\beta$. It is allowed to touch at the local point **r** in the sample.

A precursor to the above case is the situation where all states in lead $\beta$ is empty and lead $\beta$ is allowed to carry current away from the sample maximally, by quantum tunneling. In that case the coherent current from lead $\gamma$ to $\alpha$ at $T = 0K$ is given by [4] $I_{\alpha\gamma} = \frac{e}{h}|S'_{\alpha\gamma}|^2$ where $|S'_{\alpha\gamma}|$ can be related to LPDOS as

$$|S'_{\alpha\gamma}|^2 = |S_{\alpha\gamma}|^2 - 4\pi^2 |t|^2 \vartheta_\beta \rho_{lpd}(\alpha, r, \gamma) \qquad (2)$$

In Eq. (2), $|S'_{\alpha\gamma}|^2$ is transmission probability in presence of lead $\beta$ and $|S_{\alpha\gamma}|^2$ in absence of lead $\beta$. The quantity $|t|^2$, is trasition probability from $r$ to the lead $\beta$, will be zero when lead $\beta$ is not there. The quantity $\nu_\beta$ corresponds to the DOS of lead $\beta$ or number of empty states available in lead $\beta$. In absence of lead $\beta$ we will have only the first coherent term on RHS i.e., $|S_{\alpha\gamma}|^2$. The negative sign of second term indicates loss of coherent electrons into the lead $\beta$. To get Eq. (1) one has to reinject this current back into the sample as incoherent current so that lead $\beta$ does not carry any current. $\rho_{lpd}(\alpha, \mathbf{r}, \gamma), \rho_{pd}(\alpha, \gamma)$ and $\rho_d$ can be defined as [3]

$$\begin{aligned}\rho_{lpd}(\alpha, \mathbf{r}, \gamma) &= -\frac{1}{2\pi}|S'_{\alpha\gamma}|^2 \frac{\delta\theta_{s\alpha\gamma}}{\delta U(r)}, \\ \rho_{pd}(\alpha, \gamma) &= \int \rho_{lpd}(\alpha, \mathbf{r}, \gamma) d^3\mathbf{r}, \\ \rho_d &= \sum_{\alpha\gamma}\int \rho_{lpd}(\alpha, \mathbf{r}, \gamma) d^3\mathbf{r} \end{aligned} \qquad (3)$$

Where $\frac{\delta}{\delta U(r)}$ is functional derivative. $\rho_{lpd}(\alpha, r, \gamma)$, is a novel local object in QM, derived from the idea of Larmor precession time ($\tau_{lpt}$) [3]. It gives the local density of states at $\mathbf{r}$ for those partial electrons that are going from $\gamma$ to $\alpha$ via $\mathbf{r}$. Eq. (2) claims the reduction in coherent current due to lead $\beta$ is precisely determined by this local object at the point $\mathbf{r}$. In conventional QM such an object does not exist theoretically or practically. We will make a first principle calculation to show its theoretical existence, even that being a novel contribution to theory of QM. According to the concepts of QM, $|S'_{\alpha\gamma}|^2$ and $|S_{\alpha\gamma}|^2$ cannot be related by any local entity. Equations (2) and (4) claim otherwise. We therefore consider a simple version of the setup in Fig.1 in Fig.2. The sample is the three prong potential shown in bold lines. A picture of three prong potential is depicted in Fig. 2. $V_1$ is the tunneling barrier which allows or forbids tunneling out of electrons to lead 2 that acts as lead $\beta$. The tunneling barrier $V_1$ can be increased to cut off lead $\beta$ theoretically to get $|t_{31}|^2$ and then adjusted to give $|t'_{31}|^2$. These quantities can be calculated using standard quantum mechanics. The point $r$ is thus the tip of region VI.

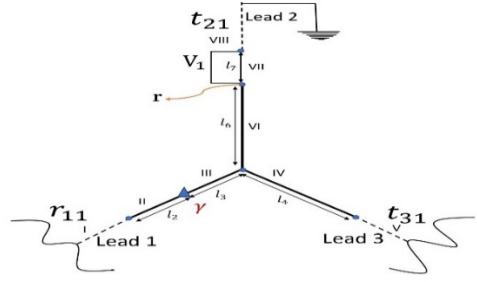

Figure 2. Three-pronged potential is shown with different regions like I, II, III, IV, V, VI, VII and VIII with the lengths like $l_1$, $l_2, l_3, l_4, l_5, l_6, l_7$ and $l_8$, respectively. Left lead, right lead and middle lead are denoted as $\alpha = 1, \gamma = 3$ and $\beta = 2$. V is the potential in dark solid lines. In Fig., $r_{11}$ is reflection amplitude from lead 1 to 1, $t_{31}$ is transmission amplitude from lead 1 to 3 and $t_{21}$ is transmission amplitude from lead 1 to 2.

After simplifying Eq.2 for our system described in Fig.2, one can find
$$|t'_{31}|^2 - |t_{31}|^2 = |t'_{31}|^2(\theta'_{t_{31}} - \theta_{t_{31}}) \tag{4}$$
We will numerically verify Eq.4 in the next section.

## RESULTS AND DISCUSSIONS

Any change in the system will change $t', t, \theta'$ and $\theta$ and in general they will not satisfy Eq. (4). One can always test this. But only for the specific changes as described with Fig.2. We can expect agreement between LHS and RHS of Eq. (4). This is plotted in Fig.3. For further details see figure caption. Normally we always expect $|t'|^2$ to be less than $|t|^2$ but here it oscillates. Note that $|t'_{31}|^2 - |t_{31}|^2$ is becoming very systematically positive and negative, and it agrees in amplitude and frequency the RHS of Eq.4.

Although the two curves are the same in amplitude and frequency, one can find a phase difference between the two. Because of tunneling, lead 2 slightly disturbs the quantum states at the local point r of the scatterer. Those electrons which tunneling out to lead 2 have some energy dependent magnitude and phase. We believe that this will motivate experimentalists to verify the same for a realistic system as shown in Fig.1.

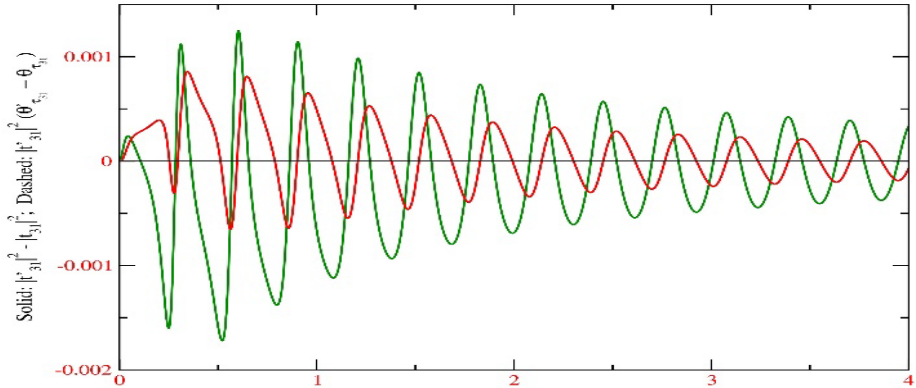

Figure 3. Plot of difference in transmission $|t'_{31}|^2 - |t_{31}|^2$ (solid curve) and difference in phase shift $\theta'_{t_{31}} - \theta_{t_{31}}$ (dashed curve) in presence of lead 2 in Fig.2 as a function of energy or wave vector k. We have taken $l_2 = l_{2n} = 0$, $l_3 = l_{3n} = 0$, $l_4 = l_{4n} = 0$, $l_6 = l_{6n} = 10$, $l_7 = l_{7n} = 0.001$, $\gamma = \gamma_n = 0$, $V_1 = 1000$, $V_{1n} = 995$, and $V = V_n = 0$ in all three arms of Fig. 2. With $\delta V1 = 5$ results in $t \rightarrow t'$ and $\theta \rightarrow \theta'$.

## DISCRETE METHOD OR TBM

A discrete version [5] of the three-prong potential which gives an easy way to deal with the complicated mesoscopic system is schematically shown in Fig. 6 and see details in Fig. caption. One can write Schrodinger equations for each site [5] and thus one can eliminate site 5 to get $1D$ chain with effective site energy at 2 i.e., $\epsilon_2'$ where

$$\epsilon_2' = \epsilon_2 + \frac{V^2}{E-\epsilon_5} \ ; \ \epsilon_2 \neq 0 \ \& \ \epsilon_5 = 0 \ ; \ \epsilon_2 = 0 \ \& \ \epsilon_5 \neq 0 \tag{5}$$

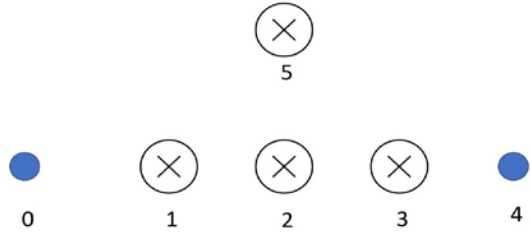

Figure 4. . A discrete version of three prong potential where the left and right most solid sites can be seen as the left and right most solid sites can be seen as left and right reservoirs of Fig. 2. Sites 1 and 3 are left lead ($\alpha = 1$) and right lead ($\gamma = 3$). Site 2 and 5 ($\beta = 2$) form a dimer that scatters. The site energies for sites 0, 1, 2, 3, 4, and 5 are $\varepsilon_0, \varepsilon_1, \varepsilon_2, \varepsilon_3, \varepsilon_4$ and $\varepsilon_5$, respectively.

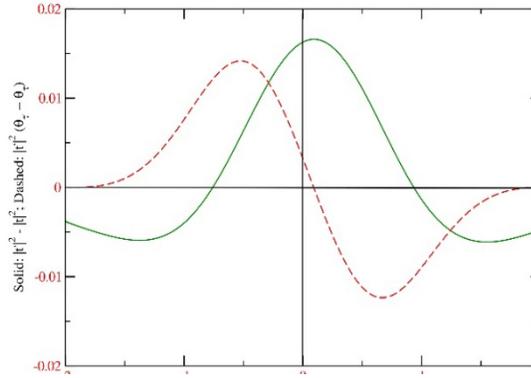

Figure 5. Plot of transmission $|t|^2$ (solid curve) and phase shift $\theta_t$ (dashed curve) as a function of incident energy E with $\epsilon_2 = 0$ and $\epsilon_5 = 0.07$ I. In TBM with imaginary energy of site 5 means it is modeled as an STM tip which absorbs electrons like lead β in the continuum model. The STM tip can not be put physically in discrete model.

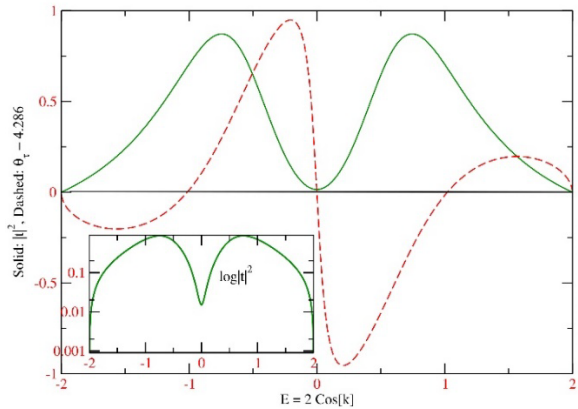

Figure 6. Solid curve is for changed transmission $|t'|^2$ and dashed curve is for changed phase shift $|t'|^2$ ($\theta_{t'} - \theta_t$). The site energies are taken as $\epsilon_5 = -0.1 + 1.5$ I and $\epsilon_{5n} = -0.2 + 1.5$ I with $\epsilon_2 = \epsilon_{2n} = 0$.

Where, $\epsilon_2$ and $\epsilon_5$ are site energies at 2 and 5, respectively. V is the hopping parameter. E is the incident energy of the electron given as $E = 2\, Cos(ka)$, where a is lattice spacing. We can see that, in Fig.5, there is gradual phase drop in phase shift $\theta_t$ that indicates negative local partial states. In Fig.6 we show agreement between LHS and RHS of Eq.4 for the discrete system of Fig.4.

## CONCLUSIONS

For the system in Fig.2 and Fig.4 validity of Eq.4 and hence Eq.2 is established. It proves that there is a local object $\rho_{lpd}(\alpha, r, \gamma)$ in QM that determine mesoscopic currents and hence response. Tight binding model can be also used to study the manifestations of negative LPDOS. Both methods show that $\rho_{lpd}$ can become negative as well as positive. Negativity of $\rho_{lpd}$ is thus experimentally detectable in similar setups as in Fig.2 and Fig.1. It implies that negative $\rho_{lpd}$ results in coherent electrons drawn into the system rather than losing them because of lead $\beta$.